\documentclass[12pt, secnumarabic, amssymb, nofootinbib, aps, prd]{revtex4}

\usepackage[english]{babel}
\makeatletter\AtBeginDocument{\let\@elt\relax}\makeatother 
\usepackage{amsmath}
\usepackage{amssymb}
\usepackage{amsbsy}
\usepackage{amstext}
\usepackage{graphicx}
\usepackage{hyperref}
\usepackage{pst-node}
\usepackage{verbatim}
\usepackage{tikz}
\usepackage{cancel}
\usepackage{braket}
\usepackage[normalem]{ulem}
\usetikzlibrary{decorations.pathmorphing}

\newcommand{\be}{\begin{eqnarray}}
\newcommand{\ee}{\end{eqnarray}}
\newcommand{\bdm}{\begin{displaymath}}
\newcommand{\edm}{\end{displaymath}}
\newcommand{\ds}{\displaystyle}
\newcommand{\nn}{\nonumber}
\newcommand{\ba}{\begin{array}}
\newcommand{\ea}{\end{array}}
\newcommand{\pa}[1]{\left(#1\right)}
\newcommand{\paq}[1]{\left[#1\right]}

\begin{document}

\title{ Gravitational radiation contributions\\
to the two-body scattering angle}

\author{Gabriel Luz Almeida}
\email{gabriel.luz@fisica.ufrn.br}
\affiliation{Departamento de F\'\i sica Te\'orica e Experimental, Universidade Federal do Rio Grande do Norte, Avenida Senador Salgado Filho, Natal-RN 59078-970, Brazil}

\author{Stefano Foffa}
\email{stefano.foffa@unige.ch}
\affiliation{D\'epartement de Physique Th\'eorique and Gravitational Wave Science Center, Universit\'e de Gen\`eve, CH-1211 Geneva, Switzerland}

\author{Riccardo Sturani}
\email{riccardo.sturani@unesp.br}
\affiliation{Instituto de F\'\i sica Te\'orica, UNESP- Universidade Estadual Paulista \& ICTP South American Institute for Fundamental Research, Sao Paulo 01140-070, SP, Brazil}

\begin{abstract}
We compute the contribution to the two-body scattering angle of a specific class of interactions involving the exchange of gravitational radiative degrees of freedom, including the nonlinear memory
process and square of radiation reaction effects.
Our computation is performed directly from the equations of motion,
thus computing the overall effect of both conservative and dissipative processes.
Such contributions provide in principle the last missing ingredients to compute the scattering angle at fifth post-Newtonian, at fourth post-Minkowskian order.
\end{abstract}

\maketitle

\section{Introduction}

With the detection rate of gravitational wave (GW) signals from compact binary inspirals and coalescences approaching one per week in the latest O3 LIGO-Virgo sciece run \cite{LIGOScientific:2020ibl,LIGOScientific:2021djp}, and with third generation and space detector projects already on the way \cite{Punturo:2010zza,Reitze:2019iox,LISA:2017pwj}, GWs and the analytic modeling of binary dynamics are attracting more interest then ever, a trend which is likely to continue in the foreseeable future.
It comes then with no surprise that compact binary dynamics is drawing the attention of high energy theorists which sided with the already active general relativity (GR)-oriented community,  following several indications that processes \cite{Damour:2016gwp} and methods \cite{Goldberger:2004jt,Foffa:2016rgu,Bern:2019nnu} usually associated with particle physics investigations can equally well and successfully describe classical gravitational processes.

The most commonly used approximation methods to tackle the two-body problem in gravity are the post-Newtonian (PN) \cite{Blanchet:2013haa},
post-Minkowskian (PM) \cite{Bertotti:1960wuq,Bertotti:1956pxu,Damour:2016gwp}, and the self-force (SF) schemes, see e.g.~\cite{Barack:2018yvs} for a recent review.
While the PM approximation is a perturbative expansion in the gravitational coupling $G$ only, the PN framework adopts the binary constituents' relative velocity $v$
as expansion parameter, and it mixes velocity and gravitational
self-interaction corrections by using Kepler relation $v^2\simeq GM/r$, being $M$ the binary total mass and $r$ the binary constituents' distance; finally, the SF scheme is obtained by expanding in the ratio between binary constituent masses $m_{1,2}$, $q\equiv m_1/m_2$, or rather its symmetric extension $\nu\equiv \mu/M$, with $\mu$ being the reduced mass.

Focusing on the spin-less, conservative dynamics sector, the current state of the art is provided by the next-to-leading order (NLO) in SF \cite{Detweiler:2008ft}, (NLO)$^4$ in PN (henceforth 4PN \cite{Jaranowski:2013lca,Damour:2014jta,Bernard:2017bvn,Marchand:2017pir,Foffa:2019rdf,Foffa:2019yfl,Blumlein:2020pog}), and (NLO)$^3$ in PM (3PM \cite{Bern:2019nnu,Damour:2020tta,Kalin:2020fhe}),
in addition with several partial results available at 5PN \cite{Bini:2019nra,Foffa:2019hrb,Foffa:2019eeb,Foffa:2020nqe,Blumlein:2020pyo,Blumlein:2021txe} and 6PN \cite{Bini:2020nsb,Bini:2020hmy,Bini:2020rzn,Blumlein:2020znm,Blumlein:2021txj}, at 4PM \cite{Bern:2021dqo,Bern:2021yeh,Dlapa:2021npj,Dlapa:2021vgp}, and at 2SF \cite{Bini:2016cje}.

In view of checking consistency among results obtained in different approximation schemes, 
the scattering angle in a two-body process is particularly convenient as it is gauge invariant and it encapsulates the complete two-body
dynamics. Moreover a simple heuristic argument \cite{Damour:2016gwp} predicts that the PM-expanded scattering angle has a simple $\nu$ dependence:
$n$PM expression involves at most $[(n-1)/2]$SF terms.\footnote{We denote by $[x]$ the integer part of $x$.}
The computation of the scattering angle at 3PM has been completed for both conservative \cite{Bern:2019nnu,Kalin:2020fhe} and dissipative \cite{Damour:2020tta} effects, and it satisfies
the previous argument about $\nu$ scaling, as well as the ultra-relativistic limit $m_{1,2}\to 0$ \cite{Damour:2020tta,DiVecchia:2021ndb}.
At 4PM order \cite{Bern:2021yeh, Dlapa:2021vgp} the scattering angle has been computed adopting a specific prescription for the Green's function
which projects out dissipative effects, and while such result is not expected to
reproduce the entire scattering angle at 4PM order, it does satisfy the requirement of
absence of 2SF terms.
On the other hand, the 5PN results collected so far appear to be in tension with the expected $\nu$ scaling at 4PM \cite{Bini:2021gat}, and a crucial role in
this disagreement could be played by hereditary terms \cite{Blanchet:1987wq}, tails and memory, which enter the conservative dynamics at 4PN \cite{Foffa:2011np,Galley:2015kus} and whose understanding at 5PN is not yet settled \cite{Foffa:2019eeb,Foffa:2021pkg,Blumlein:2021txe}.

Whereas tails (interaction of the GW with the quasi-static curvature generated by the binary system) can be considered well understood, even at all PN orders \cite{Blanchet:2019rjs,Almeida:2021xwn}, the same cannot be said about
non-linear memory terms (interaction of GWs among themselves); in particular it emerged that (contrarily to tails) their conservative and dissipative effects are not trivially separable \cite{Blumlein:2021txe}.

While investigations so far \cite{Bini:2021gat,Blumlein:2021txe,Bern:2021yeh,Dlapa:2021vgp} have been based on attempts to isolate the conservative contributions to the scattering angle (or, more recently,  to treat them separately from the dissipative ones \cite{Kalin:2022hph,Jakobsen:2022psy}),
in the present work we tackle the problem by working exclusively at the level of equations of motion (henceforth, {\it eom}),
thus automatically including both conservative and dissipative effects.
As a consequence, contrarily to previous treatments, here we find the \emph{total} contribution of memory (and memory-like) terms to the 4PM scattering angle (leaving ${\cal O}(G^5)$ to further investigations), without distinction between conservative and dissipative parts. 
To obtain our result we need to use the \emph{in-in} formalism
\cite{Schwinger:1960qe,Keldysh:1964ud,Galley:2009px}, which is necessary
when dealing with Green's functions for the exchange of radiative modes, that are intrinsically
non time-symmetric, whereas the standard in-out formalism is well suited to treat processes mediated
by potential modes, and can be applied straightfowardly to processes involving at most two radiative modes, as is the case for tails \cite{Foffa:2021pkg}.
Note that all the processes we are considering here have no radiation going to infinity, hence the dissipation arises from integrating out massless modes
in internal processes.
Our {\it eom}-based approach also brings naturally to the inclusion of
effects quadratic in the radiation reaction force,
as suggested in \cite{Bini:2021gat},
which are indeed expected to start playing a role at 5PN order and are also computed in this work.
By adding our results to the other partial results obtained considering tail and potential processes at 5PN,
computed in \cite{Blumlein:2021txe, Foffa:2019eeb}, one is expected to complete the 5PN subsector of the 4PM scattering angle.
However, contrary to expectations, we find that such subsector \emph{does still contain a 2SF term}; the failure to meet the expected $\nu$ scaling calls for additional investigations to recompose the discrepancy.

The paper is organized as follows: in Section \ref{sec:method} we describe the
dynamical processes we are going to consider and sketch the derivation of the scattering angle from the {\it eom}; in Section \ref{sec:scaQQ} we outline the actual
computation, which is summarized and compared with known results in Section \ref{sec:summary}. Finally , Section \ref{sec:concl} contains our concluding remarks.

\section{Processes and procedure}\label{sec:method}
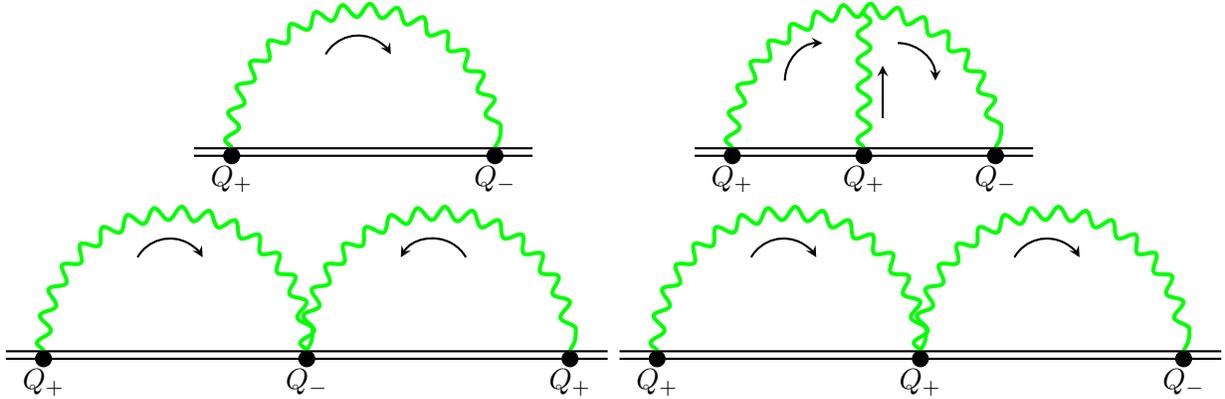
\begin{figure}
  \begin{center}
        \begin{tikzpicture}
      \draw [black, thick] (0,0) -- (4.5,0);
      \draw [black, thick] (0,-0.1) -- (4.5,-0.1);
      \draw[decorate, decoration=snake, line width=1.5pt, green] (0.5,0)  arc (180:0:1.75);
      \filldraw[black] (0.5,-0.1) circle (3pt) node[anchor=north] {$Q_+$};
      \filldraw[black] (4.,-0.1) circle (3pt) node[anchor=north] {$Q_-$};
     \draw[-stealth,thick] (1.75,1.25)  arc (150:30:.5);
    \end{tikzpicture}\hspace{2cm}
    \begin{tikzpicture}
      \draw [black, thick] (0,0) -- (4.5,0);
      \draw [black, thick] (0,-0.1) -- (4.5,-0.1);
      \draw[decorate, decoration=snake, line width=1.5pt, green] (2.25,0) -- (2.25,1.85) ;
      \draw[decorate, decoration=snake, line width=1.5pt, green] (0.5,0)  arc (180:0:1.75);
      \filldraw[black] (0.5,-0.1) circle (3pt) node[anchor=north] {$Q_+$};
      \filldraw[black] (2.25,-0.1) circle (3pt) node[anchor=north] {$Q_+$};
      \filldraw[black] (4.,-0.1) circle (3pt) node[anchor=north] {$Q_-$};
       \draw[-stealth,thick] (2.5,.4)--(2.5,1.1);
      \draw[-stealth,thick] (1.2,.9)  arc (180:90:.5);
      \draw[-stealth,thick] (2.7,1.4)  arc (90:0:.5);
    \end{tikzpicture}\\
      \begin{tikzpicture}
      \draw [black, thick] (0,0) -- (8,0);
      \draw [black, thick] (0,-0.1) -- (8,-0.1);
      \draw[decorate, decoration=snake, line width=1.5pt, green] (0.5,0)  arc (180:0:1.75);
      \draw[decorate, decoration=snake, line width=1.5pt, green] (4.,0)  arc (180:0:1.75);
      \filldraw[black] (0.5,-0.1) circle (3pt) node[anchor=north] {$Q_+$};
      \filldraw[black] (4.,-0.1) circle (3pt) node[anchor=north] {$Q_-$};
      \filldraw[black] (7.5,-0.1) circle (3pt) node[anchor=north] {$Q_+$};
           \draw[-stealth,thick] (1.75,1.25)  arc (150:30:.5);
           \draw[stealth-,thick] (5.25,1.25)  arc (150:30:.5);
    \end{tikzpicture}
    \begin{tikzpicture}
      \draw [black, thick] (0,0) -- (8,0);
      \draw [black, thick] (0,-0.1) -- (8,-0.1);
      \draw[decorate, decoration=snake, line width=1.5pt, green] (0.5,0)  arc (180:0:1.75);
      \draw[decorate, decoration=snake, line width=1.5pt, green] (4.,0)  arc (180:0:1.75);
      \filldraw[black] (0.5,-0.1) circle (3pt) node[anchor=north] {$Q_+$};
      \filldraw[black] (4.,-0.1) circle (3pt) node[anchor=north] {$Q_+$};
      \filldraw[black] (7.5,-0.1) circle (3pt) node[anchor=north] {$Q_-$};
           \draw[-stealth,thick] (1.75,1.25)  arc (150:30:.5);
           \draw[-stealth,thick] (5.25,1.25)  arc (150:30:.5);
    \end{tikzpicture}
    \caption{Processes giving rise to memory-like contributions in the 5PN equations of motion. Arrows indicate the orientation of the retarded propagators used in the in-in formalism. Top line: simple emission and memory; bottom line: ``double self-energy" diagrams involving one nonlinear GW-quadrupole coupling.}
    \label{fig:QQQ}
  \end{center}
\end{figure}

We study the dynamical effects of the processes shown in Figure \ref{fig:QQQ},
where wavy lines represent radiative modes emitted and absorbed
by the same source, a composite object with multipolar coupling to gravity representing the inspiralling binary system.
We work within the NRGR theory \cite{Goldberger:2004jt}, in which the fundamental coupling of point particles to gravity is traded for the equivalent theory
of a point particle coupled to potential modes, plus multipole moments coupled to radiative modes.

Radiative modes require the use of non time-symmetric Green's functions, which are incompatible with the standard in-out formalism usually
adopted in particle physics, and requires to be treated in the in-in formalism.\footnote{In previous works \cite{Galley:2015kus, Foffa:2019eeb,Foffa:2021pkg,Kalin:2022hph}, it has been shown that actually a well defined conservative Lagrangian in the standard in-out formalism can be obtained for the effective action computations when at most two radiative modes are involved, like tails.
However at 5PN the contributions of memory and radiation-reaction squared processes,
which involve three radiative modes, are intrinsically non time-symmetric.}

Integrating out the processes described in Figure \ref{fig:QQQ} results into a
generalized action functional ${\cal S}\paq{{\bf r}_+, {\bf r}_-}$, where the
$\pm$ variables are combination of the two copies of the initial physical variable.
The physical {\it eom} are recovered by deriving the generalized action functional with respect to the ``$-$'' variable and then setting the ``$+$'' ones
to the physical ones and the ``$-$'' ones to zero: $\left.\pa{\delta {\cal S}}/\pa{\delta {\bf r}_-}\right|_{{\bf r}_-=0,{\bf r}_+={\bf r}}$.
The quadrupole emission/absorption diagram (upper left in Figure \ref{fig:QQQ}) describes the radiation-reaction process (for details, see also \cite{Galley:2015kus}):
\be
{\cal S}_{rr}=-\frac{G}5\int_t Q_-^{kj}Q_{+\,kj}^{(5)}\quad\Rightarrow\quad {\bf a}^i_{rr}=-\frac{G}{5\mu}\left.\frac{\delta{Q_{-}^{kj}}}{\delta {\bf r}_-^i}Q_{+\,kj}^{(5)}\right|_{{\bf r}_-=0\,,{\bf r}_+={\bf r}}\,,
\ee
where at leading order
\be
\label{eq:Qpm}
Q_{-}^{kj}\simeq\mu \pa{{\bf r}^k_+{\bf r}^j_-+{\bf r}^j_+{\bf r}^k_--\frac23 \delta^{kj}{\bf r}_+\cdot{\bf r}_-}\,,\quad
Q_{+}^{kj}\simeq \mu \pa{{\bf r}^k_+{\bf r}^j_+-\frac13 \delta^{kj}r_+^2}\,,
\ee

from which the 2.5PN Burke-Thorne \cite{Burke:1970dnm} radiation-reaction force is derived.

The other diagrams in Figure \ref{fig:QQQ} have been first evaluated in \cite{Blumlein:2021txe} . We have recomputed them here and found the same value for the memory diagram (upper right in Figure \ref{fig:QQQ}),
\be
\label{eq:actmem}
{\cal S}_{mem}=G^2\int_t\paq{\frac15Q_{+ij}^{(4)}Q_{+jk}^{(4)}Q_{-}^{ik}-\frac25Q_{+ij}^{(4)}Q_{+jk}\pa{Q_{-}^{ik}}^{(4)}+\frac8{35}\dddot{Q}_{+ij}\dddot{Q}_{+jk}\ddot{Q}_{-}^{ik}-\frac{12}{35}\dddot{Q}_{+ij}\ddot{Q}_{+jk}\dddot{Q}_{-}^{ik}}\,,
\ee
and for the ``double self-energy" (henceforth, ``ds-e'') diagrams,
which are the ones in the second line of Figure \ref{fig:QQQ}\footnote{More precisely, for "ds-e" we agree with the value contained in the Erratum of \cite{Blumlein:2021txe}, also later confirmed in \cite{Brunello:2022zui}. Notice also that in \cite{Blumlein:2021txe} a different normalization for the $\pm$ variables is adopted with respect to (\ref{eq:Qpm}), or \cite{Galley:2009px}, resulting in the appearence of spourious extra $\sqrt{2}$ factors. }
\be\label{eq:acttits}
{\cal S}_{ds-e}=G^2\int_t\paq{-\frac12Q_{+ij}^{(4)}Q_{+jk}^{(4)}Q_{-}^{ik}+Q_{+ij}^{(4)}Q_{+jk}\pa{Q_{-}^{ik}}^{(4)}}\,.
\ee

The aim of this paper is to compute the contribution to the scattering angle $\chi$ as
\be
\chi ={\rm arcos}\pa{\frac{{\bf p}^+\cdot {\bf p}^{-}}{|{\bf p}^+||{\bf p}^-|}}\,,
\ee
where ${\bf p}^{\pm}$ is the momentum of the relative motion at
$t=\pm\infty$, with \be\label{eq:chiofa}
{\bf p}^+={\bf p}^-+\mu \int_{-\infty}^{\infty}{\rm d} t\ {\bf a}\,.
\ee
More specifically, we will compute 4PM, 5PN contributions to the scattering angle of the processes depicted in Figure \ref{fig:QQQ}.

\section{Computation of the scattering angle}\label{sec:scaQQ}
We apply here the {\it eom}-centered procedure introduced in the previous
section to the case of memory-like contributions to the scattering angle.
For the sake of generality, we consider action functionals of the form
\be
{\cal S}_{Q^3}&=& G^2\sum_{n=3,4} \pa{A_n Q^{-++}_n+B_n Q^{++-}_n}\,,\quad {\rm where}\quad Q^{\pm\pm\pm}_n\equiv \int_t {\rm Tr}\paq{Q_{\pm}^{(n)}Q_{\pm}^{(n)}Q_{\pm}^{(8-2n)}}\nn
\ee
represent the structures introduced in Equations (\ref{eq:actmem},\ref{eq:acttits}).
This gives the following \emph{eom} at leading order
\be
{\bf a}_{Q^2}^k&=&G^2 \pa{{\bf r}^i\delta_{kl}+{\bf r}^l\delta_{ki}-\frac{2\delta_{il}}3{\bf r}^k} \sum_{n=0}^4\alpha_{8-n}{Q_{ij}}^{(8-n)} Q_{jl}^{(n)}\,,\\
&&\alpha_8=A_4\,,\quad \alpha_7=4A_4\,,\quad\alpha_6=6A_4 -A_3\,,\nn\\
&&\alpha_5=4A_4 -4A_3+ 2B_3\,,\quad\alpha_4=A_4+B_4 -3A_3+ 2B_3\nn\,.
\ee

When applying the order reduction of time derivatives by replacing systematically all the accelerations with their Newtonian value, one finds that in general
the lowest nonvanishing contribution is $G^3$, but it turns out that such
terms do not contribute to the scattering
  angle and they can be eliminated
from the equations of motion via a suitable variable change ${\bf r}\rightarrow{\bf r} +\delta{\bf r}$, with ${\rm lim}_{t\rightarrow\pm\infty}\delta{\bf r}=0$ so that the scattering angle is not affected.
One is then left with
\be
\label{eq:aQQG4}
{\bf a}_{Q^2}&=&\frac{G^4 M^4 \nu^2}{r^6}\paq{\pa{c_1 v^4+c_2 v^2 v_n^2 +c_3 v_n^4} {\bf r}+\pa{c_4 v^2 +c_5 v_n^2}v_r{\bf v}}+{\cal O}\pa{G^5}\,,
\ee
where
\be
\label{eq:ci}
&&c_1=-\frac{376 A_3+692 A_4+736 B_3-208 B_4}9\,,\quad c_2=\frac{688 A_3-148A_4+2384 B_3+208 B_4}3\,,\nn\\
&&c_3=-\frac{736 A_3-2072 A_4+2432 B_3+336 B_4}3\,,\ c_4=-\frac{176 A_3-284A_4+320 B_3+40 B_4}3\,,\\
&&c_5=\frac{488 A_3-836 A_4+720 B_3+120 B_4}3\nn\,,
\ee
$v_r\equiv {\bf v}\cdot {\bf r}$ and $v_n\equiv v_r/r$.
We stress again that no distinction is attempted here between conservative
and nonconservative terms, as in general the action functionals cubic in
nonconserved multipoles, like ${\mathcal S}_{mem}$ and ${\mathcal S}_{ds-e}$,
contain both, as explicitly shown in \cite{Blumlein:2021txe}.
This is also reflected by the impossibility of writing the mechanical energy
loss ${\bf a}_{Q^2}\cdot{\bf v}$ as a total derivative of contractions of three
generic $Q_{ij}(t)$'s. However we incidentally note that at 5PN order,
where we are
allowed to use the leading order expression for $Q_{ij}$ and
substitute the acceleration with their Newtonian expression,
this is no longer true and
${\bf a}_{Q^2}\cdot{\bf v}$ is indeed a total derivative, meaning that the
associated GW flux is actually entirely given by a Schott term.

Before showing the results for the scattering angle, there are other contributions of the same kind to be taken into account. This can be understood by writing the {\it eom} as
\be
{\bf a}=-\frac{G M}{r^3} {\bf r}+\dots+{\bf a}_{rr}+\dots+\left.{\bf a}_{Q^2}\right|_{mem,d s-e}+\dots\,,
\ee   
where the dots represent all other known ${\cal O}(nPN)$ conservative terms ($n\leq 5$),
the ${\cal O}(3.5PN)$ radiation reactions ones, as well as still unknown contributions beyond 5PN.

As ${\bf a}_{Q^2}$ is of 5PN order, for consistency one must include,
besides all other known 5PN contributions reported in \cite{Foffa:2019eeb} and
\cite{Blumlein:2020pyo}, the effect due to the 2.5PN Burke-Thorne
radiation-reaction acceleration
\be
{\bf a}_{rr}^i \simeq{\bf a}_{BT}^i=-\frac 25 G Q^{(5)}_{ij}x^j\,,
\ee 
when, in the $G$-order reduction, the accelerations in the r.h.s. are replaced by
\be
{\bf a}\rightarrow-\frac{G M}{r^3}{\bf r}+{\bf a}_{BT}\,; 
\ee
this generates ultimately a 2.5PN part of ${\bf a}_{BT}$
\be
\label{eq:aBT_harm}
\ba{rl}
\ds{\bf a}_{BT-2.5PN}=&\ds-\frac{G^2 M^2 \nu}{r^4}\left[\pa{40 v_n^2-\frac{144}5 v^2}v_n{\bf r}+\pa{\frac{48}5v^2-24 v_n^2}r\ {\bf v}\right.\\
  &\ds \qquad\left.-\frac{16}{5}\frac{GM}{r}\pa{r{\bf v}+\frac13 v_n {\bf r}}\right]
\,,
\ea
\ee
and a 5PN part, ${\bf a}_{BT^2}$, which has the same structure of ${\bf a}_{Q^2}$, and in particular can also be put in the form reported in Equation (\ref{eq:aQQG4})
with
$\pa{c_1\,\dots\,,c_5}=\pa{-\frac{5696}{225},\frac{512}{15},\frac{17344}{75},\frac{2944}{25},-\frac{5632}{25}}$.

Note that the Burke-Thorne force is responsible
for a time-odd contribution to $\chi$ at ${\cal O}(G^3)$ order,
denoted by $\chi^{rad}$, whose value at leading order in $v$ can be computed
by evaluating eq.(\ref{eq:chiofa}) perturbatively and integrating
${\bf a}_{BT-2.5PN}$ along a straight trajectory with initial relative velocity
$v_-$ and impact parameter $b$. The result
\be
\label{eq:chirad}
\chi^{rad}\simeq \frac{16\nu}{5 v_-}\pa{\frac{GM}{b}}^3
\ee
is in agreement with the one reported in Equation (7.3) of \cite{Damour:2020tta}.

Following the same strategy it is straightforward to derive the 4PM and 5PN
contribution to the scattering angle associated to ${\bf a}_{Q^2}$ and ${\bf a}_{BT^2}$.
Using the same notation as in \cite{Bini:2021gat}, that is
\be
\frac12 \chi=\sum_{n\geq 1}\frac{\chi_n(p_{\infty}\,,\nu)}{j^n}\,,\quad
p_{\infty}=\frac{v_-}{\sqrt{1-v^2_-}}\,,\quad j=\frac{J}{G m_1 m_2}\,,
\ee
being $J$ the incoming angular momentum, we find the following contribution to the 5PN (that is ${\cal O}(p_{\infty}^6)$) part of $\chi_4$ (4PM):
\be\label{eq:chi4}
\chi_4^{Q^2, BT^2}=\frac12 \alpha \pi p_{\infty}^6\,,\quad{\rm with}\quad \alpha=-\frac{48 c_1 + 8 c_2 + 3 c_3}{128}\nu^2\,.
\ee
Substituting the values of $c_i$ from (\ref{eq:ci}), the numerical coefficient
$\alpha$ in Equation (\ref{eq:chi4}) evaluates to
$\pa{\frac{85}{12}A_3+\frac{755}{48}A_4-\frac{83}8B_4}\nu^2$ for ${\bf a}_{Q^2}$,
that is $\alpha_{mem}=-\frac{2267}{210}\nu^2$ for the memory and $\alpha_{de-e}=\frac{251}{12}\nu^2$ for the double self-energy\footnote{We remind that $A_{3,4},B_{3,4}$ can be read from Equations
  (\ref{eq:actmem},\ref{eq:acttits}), their values
being $A_3=-\frac{12}{35}$, $A_4=-\frac 25$, $B_3=\frac 8{35}$, $B_4=\frac 15$ for the memory, and $A_3=B_3=0$, $A_4=1$, $B_4=-\frac 12$ for the double self-energy.}, and to $\alpha_{BT^2}=\frac{97}{50}\nu^2$ for the 5PN part of the Burke-Thorne acceleration.

So far we have evaluated the scattering angle given by 5PN, ${\cal O}(G^4)$, acceleration terms by computing it
on straight lines trajectories, as discussed at the end of Section \ref{sec:method}. Besides this we need to add the contribution obtained by evaluating the
2.5PN acceleration ${\bf a}_{BT-2.5PN}$ on the $2.5$PN trajectory, which
can be computed by integrating
\be\label{eq:modtraj}
\frac{{\rm d}^2{\bf r}_{BT-2.5PN}}{{\rm d}t^2}\simeq -\frac{G^2 M^2 \nu}{r_0^4}\paq{\pa{40 \pa{v_n}_0^2-\frac{144}5 v_0^2}{v_n}_0{\bf r}_0+\pa{\frac{48}5v_0^2 -24 {v_n}_0^2}r_0{\bf v}_0}\,.
\ee

Following a procedure also discussed in \cite{Kalin:2020mvi}, the evaluation
of Equation (\ref{eq:chiofa}), with ${\bf a}={\bf a}_{BT-2.5PN}$, truncated at
${\cal O}(G^2)$, on the trajectory ${\bf r}_{BT-2.5PN}(t)$ obtained from
Equation (\ref{eq:modtraj}).
gives the further contribution to the scattering angle denoted as
$\alpha_{BT-{\bf r}_{BT}}=\frac{479}{25}\nu^2$, which should be combined
with $\alpha_{BT^2}$ to give the total radiation-reaction-squared contribtion.

\section{Summary}
\label{sec:summary}
It is useful at this point to summarize our findings and to integrate them with
other published results.
With reference to Equation (\ref{eq:chi4}), we have determined the following contributions to $\alpha$:
\be
\ba{l}
\ds \alpha_{mem}=-\frac{2267}{210}\nu^2\,,\quad\alpha_{ds-e}=\frac{251}{12}\nu^2\,,\\
\ds \alpha_{RR^2}\equiv\alpha_{BT^2}+\alpha_{BT-{\bf r}_{BT}}=\pa{\frac{97}{50}+\frac{479}{25}}\nu^2=\frac{211}{10}\nu^2\,.
\ea
\ee

These have to be added to other contributions of the same order already known in the literature, that is the ones coming from potential modes and tail interactions \cite{Foffa:2019hrb,Blumlein:2020pyo,Foffa:2019eeb};
still using the same notation as \cite{Bini:2021gat}, we can write the ``energy rescaled scattering coefficient'' $\tilde \chi_4\equiv h^3 \chi_4$, with
$h^2=1+2\nu(\sqrt{1+p_\infty^2}-1)$, as
\be
\tilde{\chi}_4-\chi^{Schw}_4&=&\pi\left\{-\frac{15}4\nu+\pa{\frac{123}{256}\pi^2-\frac{557}{16}}\nu p_\infty^2+\paq{-\frac{6113}{96}+\frac{33601}{16384}\pi^2-\frac{37}{5}\log\pa{\frac{p_\infty}2}}\nu p_\infty^4\right.\nn\\
&&\quad\quad+\left[\pa{\frac{93031}{32768}\pi^2
  -\frac{7437721}{188160}
  -\frac{1357}{280}\log\pa{\frac{p_\infty}2}}\nu+\frac{230281}{9800}\nu^2\right.\nn\\
&&\quad\quad\quad\quad\left.  \left. +\frac12\pa{\alpha_{MO^2}+\alpha_{MJ^2}+\alpha_{LQ^2}+\alpha_{mem}+\alpha_{ds-e}+\alpha_{RR^2}}\right]p_\infty^6\right\}\,,
\ee
where we have isolated in the last line all the far zone contributions which are genuinely 5PN at leading order (and whose contribution to $\chi_4$ and $\tilde{\chi}_4$ are consequently identical). In particular, besides the previously defined coefficients, we have also the purely conservative terms $\alpha_{MO^2}=\frac{69577}{4900}\nu(1-4\nu)$, $\alpha_{MJ^2}=\frac{147}{200}\nu(1-4\nu)$, $\alpha_{LQ^2}=\frac{138}{5}\nu^2$ which are, respectively,
the contributions from mass octupole tail, magnetic quadrupole tail, and angular momentum ``failed'' tail \cite{Foffa:2019eeb,Almeida:2021xwn}.\footnote{Notice that the quantity corresponding to $\alpha_{LQ^2}$ in \cite{Bini:2021gat}, that is $\frac{207}4C_{QQL}\nu^2$, has two opposite signs in the published PRD version and in the most recent arXiv version(v4); we agree with the latter, which is also the most recent one.}
The first two lines contain the combined effect of potential modes and of the
4PN mass quadrupole tail, as well as the logarithmic term associated with all the tails.

By inserting the numerical values one finds
\be
\label{eq:chi4us}
\tilde{\chi}_4-\chi^{Schw}_4&=&\pi\left\{-\frac{15}4\nu+\pa{\frac{123}{256}\pi^2-\frac{557}{16}}\nu p_\infty^2+\paq{-\frac{6113}{96}+\frac{33601}{16384}\pi^2-\frac{37}{5}\log\pa{\frac{p_\infty}2}}\nu p_\infty^4\right.\nn\\
&&\quad\quad\left.+\paq{\pa{-\frac{615581}{19200}+\frac{93031}{32768}\pi^2-\frac{1357}{280}\log\pa{\frac{p_\infty}2}}\nu+\frac{576}{25}\nu^2} p_\infty^6\right\}\,,
\ee
which is in contradiction with the general scaling argument that constrains the $\nu$-dependence of the scattering angle to be linear at ${\cal O}\pa{G^4}$.
Notice that the approach presented in \cite{Blumlein:2021txe}, which consists in extracting a conservative part only from the memory and ds-e terms, is equivalent to setting
$\alpha_{Q^2}=\pa{\frac{85}{24}\pa{A_3+B_3}+\frac{253}{48}\pa{A_4+B_4}}\nu^2$
instead of the value reported below Equation (\ref{eq:chi4}). This translates into $\alpha_{mem}=-\frac{817}{560}\nu^2$ and $\alpha_{ds-e}=\frac{253}{96}\nu^2$,
values which are also inconsistent with the scaling
argument for the scattering angle, with or without the addition of the radiation-reaction squared contribution $\alpha_{RR^2}$. 

As a consequence of the $O(\nu^2)$ term in eq.~(\ref{eq:chi4us}),
  the quantity ${\cal M}_4^{\rm radgrav,\ finite}$ defined in  \cite{Bini:2021gat} and computed at higher order in \cite{Bern:2021yeh}, receives a problematic
  $O(\nu)$ contribution which breaks the expected mass-polynomiality.
We stress again that our eq.~(\ref{eq:chi4us}) contains contributions that have
not been considered in  \cite{Blumlein:2021txe}, \cite{Bini:2021gat}, nor in \cite{Bern:2021yeh}.

\section{Conclusions}\label{sec:concl}
We have adopted an approach based entirely on the equations of motion to compute the contribution to the scattering angle of processes
which involve nonlinear interactions of quadrupole GW radiation: nonlinear memory, quadratic emission, and second-order radiation reaction. Such processes are
characterized by the fact that conservative and nonconservative effects are mixed and not unambiguously separable, the use of the equations of motion
allows us to deal with them in an unified way. 

When added to the other already known 4PM-5PN terms, the scattering angle value found in the present work is still at odds with the expected $\nu$ dependence at ${\cal O}(G^4)$; this  means that this problem is still open, and we do not attempt here any further speculations about the origin and the persistence of the mismatch. We are however confident that the new informations contained in this work can contribute to shed some light upon this issue in the near future.

\section*{Acknowledgements}
S.F. and R.S. warmly thank all the participants in the program \emph{High Precision Gravitational Waves}, held in the spring of 2022 at the Kavli Institute for Theoretical Physics in Santa Barbara, for stimulating discussions and interactions. This research was supported in part by the National Science Foundation under Grant No. NSF PHY-1748958.
The work of R.S. is partly supported by CNPq by Grant No. 310165/2021-0.
R.S. would like to thank ICTP-SAIFR FAPESP Grant No. 2022/06350-2.
The work of G.L.A. is financed in part by the Coordena\c{c}\~{a}o de Aperfei\c{c}oamento de Pessoal de N\'{i}vel Superior - Brasil (CAPES) - Finance Code 001.
G.L.A. wish to thank INFN Padova, Geneva University and ICTP Trieste for kind hospitality and support during the completion of this work.
S.F. is supported by the Fonds National Suisse, grant $200020\_191957$, and by the SwissMap National Center for Comptence in Research.


\end{document}